\documentclass[aps,pre,twocolumn,showpacs,amssymb]{revtex4}

\usepackage[dvips]{graphicx}
\usepackage{dcolumn}
\usepackage{bm}
\def\be{\begin{equation}}
\def\en{\end{equation}}

\def\p{\partial}

\newcommand{\bi}[1]{\mbox{\boldmath$#1$}}
\def\p{\partial}
\def\bea{\begin{eqnarray}}
\def\ena{\end{eqnarray}}
\newcommand{\ppp}[3]{{\bigg(}\frac{\partial {#1}}{\partial {#2}}{\bigg )}_{#3}}
\renewcommand{\theequation}{\arabic{section}.\arabic{equation}}

\newcommand{\dis}[1]{[{#1}]}

\begin{document}
\title{ Bubble and droplet motion in binary mixtures:\\ 
Evaporation-condensation mechanism  and Marangoni effect    
}  
\author{Akira Onuki}
\affiliation{Department of Physics, Kyoto University, Kyoto 606-8502,
Japan}



\date{\today}

\begin{abstract}

Bubble and droplet 
motion in  binary mixtures is  studied  
in  weak heat and diffusion fluxes 
and in gravity   by 
solving the linearized hydrodynamic equations 
supplemented with appropriate surface boundary 
conditions.   Without gravity, the velocity field 
is induced by  evaporation and condensation  
at the interface and  by the Marangoni effect due to 
a surface tension gradient. 
In pure fluids, the latter nearly vanishes   
since the interface temperature tends to the coexistence 
temperature $T_{\rm cx}(p)$ even in heat flow. 
In binary  mixtures, the velocity field 
can be much enhanced  by the Marangoni effect 
above a crossover concentration $c^*$  inversely proportional to 
the radius $R$ of the bubble or droplet. 
Here $c^*$  is usually very small for large  $R$ 
for  non-azeotropic mixtures.  The temperature and concentration 
deviations are also calculated. 
\end{abstract}


\maketitle


\section{Introduction}
\setcounter{equation}{0}

On earth,  
bubble motion in liquid is caused by 
gravity.   Buoyancy  effects increase  dramatically 
with increasing the droplet radius $R$. 
Due to the viscosity of 
liquid,  it moves at 
 a constant velocity $v_g$ 
estimated as \cite{Levich,Ha,Ry}
\be 
v_g\sim \frac{\rho-\rho'}{\eta}R^2g,
\en  
where $g$ is the gravity acceleration. 
Hereafter   $\rho$ and  $\eta$ 
 ($\rho'$ and $\eta'$) 
are the mass density and  the shear  viscosity 
outside (inside) the bubble. 
Another method of inducing 
bubble motion is to apply   a heat flux $Q$. 
It is well-known that a  surface 
tension variation on the surface gives rise to a  
Marangoni velocity field \cite{Levich}, causing 
 bubble motion   to  lower 
the surface free energy.   Neglecting 
phase transition, Young  {\it et al.} 
\cite{Young} calculated it  as 
\be 
v_Y \sim 
\frac{\gamma_1}{\eta\lambda}  R Q ,
\en 
where $\lambda$ is the thermal conductivity in liquid. 
The surface tension variation $\delta\gamma$ is 
assumed to be given by
\be
\delta\gamma = -\gamma_1 \delta T,
\en   
where   $\delta T$ is  the 
ambient temperature deviation, so 
$v_Y \sim \delta\gamma/\eta$.
Here $\gamma_1>0$ for  most fluids, but 
 $\gamma_1<0$ for some  fluid mixtures. 
If a liquid    
is heated from a boundary  at zero  gravity, 
 a suspended  bubble is attracted to the warmer  
boundary for $\gamma_1>0$ with a velocity  
of order  $-v_Y$,  
until it is attached to the wall.
In  heat flux on earth,  
the gravity and Marangoni mechanisms can compete.   
We mention  an experiment of applying 
heat flow from below to 
 silicone oil containing air bubbles, 
where temperature gradients  
of order $10$-$10^2$ K$/$cm 
 balanced with the buoyancy 
and held the bubbles stationary  \cite{Young}. 
Subsequent microgravity 
 experiments on the Marangoni effect 
have  been performed without phase change  
\cite{space0,Japan}.

However, 
first order phase transition 
between gas and liquid   
(evaporation and condensation) should 
take place on the bubble surface.
This is particularly the case 
for pure (one-component) fluids,  where  the pressure $p$ 
is nearly homogeneous 
outside the bubble for slow  motions and 
the interface temperature 
should then be close to the coexistence temperature  
 $T= T_{\rm cx}(p)$ at given $p$ 
 even in heat flux. Thus, in pure fluids,    
the temperature  gradient should  nearly vanish   
inside  bubbles without  Marangoni flow. 
Recently such temperature profiles have 
been calculated  from linearized hydrodynamic equations 
supplemented with appropriate surface 
 conditions \cite{OnukiPRE}
and numerically in 
the  dynamic van der Waals theory 
 \cite{OnukiPRL}.  Balance 
of a  heat flux due to 
latent heat convection and  
an  applied heat flux $Q$ gives   
 the amplitude of  the 
convective velocity inside the bubble as  
\be 
v_c \sim  \frac{Q}{\rho'T\Delta s} , 
\en   
where  $\Delta s=s'-s$ is the entropy 
difference per unit mass.
If a bubble (droplet) is suspended in liquid (gas) 
at zero gravity, it  migrates 
toward a warmer  (cooler) boundary with a  velocity  
of order $v_c$ \cite{OnukiPRL}. 
In this evaporation-condensation mechanism, 
a bubble in liquid is attracted 
to a warmer boundary wall, which is consistent with 
experiments on pure fluids without gravity   
\cite{Beysens}.

In this paper, 
we investigate bubble and droplet motion   
in  binary mixtures, where  
the Marangoni effect  and the evaporation-condensation 
can be both important.  In analyzing boiling experiments,  
Marek and Straub  \cite{St} argued that 
 convection around  a bubble 
should be dominantly caused  by  the Marangoni effect 
due to a very small amount of a  noncondensable gas. 
If a surfactant is added  as a solute, 
such a contamination effect should 
be even more enhanced \cite{Levich,Onuki1993}. 
In our theory we shall see that 
the Marangoni velocity  for   dilute 
non-azeotropic binary mixtures  is  of order,
\be 
v_M\sim 
\frac{\gamma_1 }{k_BnD_0\eta} c R  Q,
\en 
where $n$ is the number density, 
 $D_0$ is the solute 
diffusion constant in liquid, and 
$k_B$ is the Boltzmann constant. 
Balance of $v_c $ and $v_M$ yields 
a crossover concentration  $c^*$ given by 
\be
c^*=  a_1 /R, 
\en 
where  $a_1= (k_BD_0\eta/T) ( n/\rho'\Delta s |\gamma_1 |)$ is 
usually a microscopic  length far from the critical point.
See sentences below Eq.(4.16)  for $a_1$ 
near the critical point.
 Unless $\gamma_1$ is very small, $c^*$ 
is  very  small for  large droplet radius $R\gg a_1$.  
For $c\gg c^*$ 
the hydrodynamic motion 
is mostly due to  the Marangoni effect.

Furthermore,  there seems to have been no 
fundamental argument  on 
 the coefficient $\gamma_1$ in Eq.(1.3) 
in binary mixtures 
in nonequilibrium.  In this paper  we   assume 
the continuity of the temperature and the chemical potentials 
and  neglect the  pressure deviations at the interface. 
Then it follows  $\gamma_1=- (\p \gamma/\p T)_{{\rm cx},p}$, 
where the derivative is along 
the isobaric line  on the coexistence surface. 
This $\gamma_1$ tends to a well-defined limit 
in the dilute limit of binary mixtures ($c\to 0$). 
Recently,  it   has been calculated for 
 nonelectrolyte binary  mixtures \cite{OnukiJCP}. 
Particularly near the solvent criticality,  
its mean-field expression at small solute concentration  reads 
\be 
\gamma_1 =- \frac{d\gamma_0}{dT} \frac{dp_c}{dX}\frac{1}{K_{\rm Kr}},
\en 
where $\gamma_0(T)$ is the surface tension of the pure 
solvent,  $p_c(X)$ is the critical pressure dependent on  
the solute molar fraction $X$ on the critical line,   
and $K_{\rm Kr}$ is the so-called Krichevskii parameter
(having the dimension of pressure) 
 \cite{Sengers1,De,Anisimov,Shock,Russia}. See Appendix C 
for discussions of $K_{\rm Kr}$. 
While $d\gamma_0/dT<0$ for pure fluids,  
the two parameters 
$dp_c/dX$ and $K_{\rm Kr}$  
can be both positive and negative, depending on the 
solute molecular size and the solute-solvent interaction. 
 For example, if near-critical 
CO$_2$ is a solvent, 
use of  data in  Ref.\cite{Russia} 
gives the value of 
$(dp_c/dX)/K_{\rm Kr}$ in Eq.(1.7) for various solutes, 
which is  0.90 for neon  but is  $-0.81$   for pentanol  
\cite{OnukiJCP}.

In their experiment,   Vochten and   Petre 
 \cite{Petre} found that  
  the surface tension 
 between air and aqueous  mixtures containing high carbon alcohols 
more than $1$mM    
exhibits  a minimum as a function 
of the temperature at constant pressure and molar fraction.  
In such fluids, $\gamma_1<0$ 
at temperatures higher than that giving the  minimum. 
Inspired by their  finding,  consequences of negative 
$\gamma_1$  have been discussed in two-phase hydrodynamics 
particularly to develop heat pipes for utilization 
in space  \cite{Azouni,Katsumi,Abe,Savino,Savino1}. 
Remarkably, if the sign of  $\gamma_1$ 
is changed, the direction of 
the Marangoni flow 
is reversed.  As a result, if  $\gamma_1<0$, 
bubbles  are easily detached from 
the heater in boiling. This  leads to a  
liquid inflow  onto the heater 
 suppressing  its dryout,  so 
fluid mixtures with  $\gamma_1<0$ 
have been called self-rewetting fluids. 
On earth, Abe \cite{Abe} observed 
a considerable  decrease in the size of 
 rising bubbles with addition of   1-butanol (6 wt$\%$) 
 to water (where $\gamma_1<0$).  
Adding  1-heptanol (0.1 wt$\%$) 
to water, Savino {\it et al.} \cite{Savino1} 
observed bubble motion toward 
a cooler end in a horizontal glass tube.

This paper will present   linear analysis in 
 the simplest case of
 a  spherical bubble or droplet  in binary mixtures 
in weak  heat and diffusion fluxes and in gravity. 
 In Section II,  we will give linear hydrodynamic 
equations and  surface boundary 
conditions including the  Marangoni 
condition for the tangential stress \cite{Levich,Bedeaux1}.   
In Section III,  
we will solve the equations in 
steady  states in the axisymmetric geometry.  
In Section IV, 
 we will examine the 
consequences in dilute mixtures. 
Estimations  near the critical point 
will  also  be  presented.
In Section V, the velocity 
field around a bubble or droplet 
will be displayed in various cases.

\section{Basic Equations}
\setcounter{equation}{0}

\subsection{Spherical droplet}

We place  a  gas bubble in liquid with radius $R$ 
in a nonelectrolyte  binary fluid  mixture.  
We do not assume surface  adsorption 
due to the amphiphilic interaction.  
The following results can be used 
also for the case of a liquid droplet in gas by exchanging 
"liquid" and "gas". 
Suppose  an  equilibrium 
state in the  gravity-free condition 
(see Appendix A of Ref.\cite{OnukiPRE}), where 
the temperature $T$ and the chemical 
potentials of the two components $\mu_1$ and $\mu_2$ 
are  homogeneous.  
The  pressure $p$ is a constant $p_0$ in the exterior  $r>R$ 
and  is $p_0+2\gamma/R$ in the interior  $r<R$ (
from the Laplace law,  
where $\gamma$ is the 
surface tension  and $r$ is the distance 
from the bubble center.  The interior and exterior 
concentrations are determined from the 
thermodynamics of binary mixtures \cite{OnukiJCP}.

We then apply   weak heat and diffusion 
fluxes  and a gravity acceleration $g$.  
They are all along the $z$ axis taken 
to be  in the upward vertical direction $\parallel {\bi e}_z$.  
Hereafter ${\bi e}_z$ 
denotes  the unit vector along the $z$ axis. 
The gradients of the temperature $T$ and 
 the mass fraction 
$c$  are homogeneous far from the bubble  
and are written as  
\be 
{\cal T}= \bigg(\frac{d T}{dz}\bigg)_{r=\infty}, 
\quad 
{\cal C}= \bigg(\frac{d c}{dz}\bigg)_{r=\infty}.  
\en
It is convenient to introduce 
the chemical potential difference per unit mass as   
\be 
\mu= \mu_2-\mu_1,
\en 
which has a gradient given by  
\be 
{\cal M}=  \bigg(\frac{d \mu}{dz}\bigg)_{r=\infty}
= \ppp{\mu}{T}{pc} {\cal T}+ \ppp{\mu}{c}{pT}{\cal C},
\en 
Here the thermodynamic derivatives  
are taken in the outer phase in the 
 isobaric condition.

The  hydrodynamic 
equations are linearized 
with respect to $\cal T$, $\cal C$, and $g$. 
The  deviations are all proportional 
to one  of these quantities. 
After a transient relaxation, 
 the bubble  moves at a constant velocity 
 $v_D$ in the vertical $z$ axis. 
We may  then take the 
origin of the reference frame 
at the bubble center and seek 
a steady axisymmetric solution 
of the hydrodynamic equations with 
appropriate boundary conditions. 
To linear order in 
$\cal T$, $\cal C$, and  $g$, 
 the bubble  shape is  spherical,  
 as  assumed in the previous theories 
\cite{Ha,Ry,Young}.  
Deviation from sphericity occurs from  
 second orders in these quantities.

In the following calculation it is 
convenient to use the spherical coordinates ($r,\theta,\varphi)$ 
with the origin at the bubble center. 
 Using the solid angles 
$\theta$ and $\varphi$ we 
 define  the three orthogonal unit vectors, 
\bea
{\bi e}_1&=& r^{-1}{\bi r}=
(\sin\theta\cos\varphi,
\sin\theta\sin\varphi,\cos\theta),   \nonumber \\
{\bi e}_2&=& \frac{\p}{\p \theta}{\bi e}_1
=(\cos\theta\cos\varphi,
\cos\theta\sin\varphi,-\sin\theta),   \nonumber \\
{\bi e}_3&=& {\bi e}_1 \times {\bi e}_2 =
(-\sin\varphi,\cos\varphi,0).  
\ena 
The interface normal is along ${\bi e}_1$. 
The velocity field $\bi v$ around the bubble 
will be  assumed to be orthogonal 
to ${\bi e}_3$ or ${\bi e}_3\cdot{\bi v}=0$.

\subsection{Hydrodynamic equations}

The mass densities 
of the two components are written as $\rho_1$ and 
$\rho_2$.   The total mass density $\rho$ and  the mass fraction $c$ 
of the second  component  are defined by 
\be 
\rho=\rho_1+\rho_2,  \quad c=\rho_2/\rho.   
\en 
The mass fluxes  of the two components are 
$\rho_1 {\bi v}-{\bi I}$ 
and $\rho_2 {\bi v}+{\bi I}$, 
where $\bi v$ is the velocity field 
and $\bi I$ is the diffusive flux. The 
 continuity equations for $\rho $ and $\rho_1$ are  \cite{Landau}  
\bea 
\frac{\p}{\p t}\rho&=&- \nabla\cdot(\rho {\bi v}),\\
 \frac{\p}{\p t} (\rho c) &=& -\nabla\cdot(\rho c {\bi v}+ {\bi I}).
\ena 
We express  $\bi I$ in terms of the 
isothermal mutual diffusion constant $D$ 
and the thermal diffusion ration $k_T$ as   \cite{Landau} 
\be 
{\bi I}=-\rho D(\nabla c  +T^{-1} k_T \nabla T).  
\en 
The momentum density  $\rho{\bi v}$ obeys 
\be 
\frac{\p}{\p t}\rho{\bi v} 
+\nabla\cdot(\rho{\bi v}{\bi v}) = 
-\nabla p +  \nabla\cdot\tensor{\sigma}- \rho g {\bi e}_z 
\en 
where $p$ is the pressure and 
 $\tensor{\sigma}=\{\sigma_{ij}\}$ is the dissipative stress tensor 
expressed in terms of the shear and bulk viscosities 
$\eta$ and $\zeta$ as 
\be 
\sigma_{ij}= \eta (\nabla_iv_j+\nabla_jv_i) 
+ (\zeta-2\eta/3) \delta_{ij}\nabla\cdot{\bi v}, 
\en 
where $\nabla_i= \p /\p x_i$ with $x_i=x,y,$ and $z$. 
The last term in Eq.(2.8) represents the gravity acceleration 
 with ${\bi e}_z$ being the unit vector along 
the $z$ axis. 
The (total) energy density $e_T= e+\rho{\bi v}^2/2$ 
consisting of the internal energy density 
$e$ and the kinetic energy $\rho{\bi v}^2/2$ 
is governed by 
\be 
\frac{\p}{\p t}e_{\rm T}= -\nabla \cdot[(e_T+ p){\bi v}
-\tensor{\sigma}\cdot{\bi v}+{\bi q}]-\rho g v_z 
\en 
where $\bi q$ is the dissipative heat current  
expressed as  \cite{Landau} 
\be 
{\bi q}= - \lambda \nabla T +  A{\bi I}. 
\en 
where $\lambda$ is the thermal conductivity 
in the absence of diffusion flux 
and $A$ is a constant. Because of 
the symmetry of the Onsager coefficients 
(see Appendix A), there is a relation 
between   $k_T$  and  $A$ given by 
\be 
k_T=\bigg [A  -{\mu}  + 
T\ppp{\mu}{T}{c p }\bigg ] \ppp{c}{\mu}{p T}.
\en

From Eqs.(2.1) and (2.3) the heat flux and 
diffusion flux are written as 
${\bi q} \to -{\cal Q}{\bi e}_z$ 
and ${\bi I} \to -{\cal I}{\bi e}_z$ 
far from the bubble $r\gg R$, where  
\bea 
{\cal Q}&=& \lambda {\cal T} +  A{\cal I}, \\ 
{\cal I}&=&\rho D({\cal C}  +T^{-1} k_T{\cal  T}).  
\ena 
The pressure gradient tends to $-\rho g {\bi e}_z$, while 
the velocity tends to $-v_D {\bi e}_z$ because the bubble is 
at rest in our reference frame. 
In the present work we linearize the hydrodynamic 
equations for the velocity field  $\bi v$ 
and the deviations  $\delta p$,  $\delta T$, and $\delta c$  
with respect to ${\cal T}$, $\cal C$, and $g$ 
in steady states  in the bulk region $r\neq R$. 
Here we  may set $\p(\cdots)/\p t=0$ neglecting the time-dependence. 
From Eq.(2.6) it follows 
   the incompressibility condition,  
\be 
\nabla\cdot{\bi v}=0.  
\en 
Then the bulk viscosity $\zeta$ does not enter in our calculations. 
This incompressibility 
condition holds even in compressible fluids  in our  linear theory. 
Notice that we neglect the term  ${\bi v}\cdot\nabla\rho$ 
 in Eq.(2.6) since it is of order ${\cal T}g$
in the gravity-induced density stratification. 
In the bulk region $r\neq R$,  Eq.(2.7)  and (2.11) yield 
\be 
\nabla^2 \delta T=0, 
\quad \nabla^2\delta c =0, 
\en 
The momentum equation becomes 
\be 
-\nabla \delta p+ \eta\nabla^2{\bi v}-\rho g {\bi e}_z ={\bi 0}.  
\en 
The pressure deviation is defined by 
$
\delta p=p-p_0$  for $r>R$ 
and $\delta p=p-p_0-2\gamma_0/R$ for $r<R$. 
For $r \neq R$, taking  the divergence of Eq.(2.18) yields 
\be 
\nabla^2\delta p=0. 
\en 
The term  $-(\p \rho/\p z)g$  is of 
the second order for $r\neq R$ 
and is negligible in our approximation.

\subsection{Interface boundary conditions} 

Next we consider the boundary conditions at the surface 
$r=R$. In this subsection all the quantities 
are those at $r=R\pm 0$.
In the following equations 
the quantities  at $r=R-0$ (immediately inside the bubble) 
are  primed  as  
 ${\bi v}'$, $\delta T'$, 
$\delta p'\cdots$, while 
those  at $r=R+0$ (immediately outside the bubble) 
are unprimed. 
Hereafter,  for any physical quantity $\cal A$, 
the symbol, 
\be 
\dis{{\cal A}}={\cal A}-{\cal A}',
\en 
denotes 
the  discontinuity of $\cal A$ at the surface. 
For example,  the entropy difference (per unit mass) and 
the mass concentration  are written as 
\be 
[s]=s-s', \quad [c]=c-c'.
\en 
The   Gibbs-Duhem relation yields  
 $[s]/[c]= -(\p \mu /\p T)_{{\rm cx},p}$ in terms 
of the chemical potential difference $\mu$. 
Hereafter $(\p\cdots/\p \cdots)_{{\rm cx},p}$ 
is  the derivative  taken 
along  the isobaric line on the coexistence surface.

The mass conservation at the surface yields 
${\bi e}_1\cdot[\rho{\bi v}]=0$. 
It is convenient to  introduce 
the mass flux $J$  through the interface by 
\be 
J= \rho {\bi e}_1\cdot{\bi v}= 
\rho' {\bi e}_1\cdot{\bi v}', 
\en 
which arises from  conversion between gas and liquid. 
We assume the continuity of the tangential velocity, 
\be 
{\bi e}_2\cdot\dis{{\bi v}}= 
{\bi e}_2\cdot{\bi v}-{\bi e}_2\cdot{\bi v}'={\bi 0}.  
\en 
The stress balance at the interface yields 
\bea
&&\dis{\delta p - {\bi e}_1\cdot\tensor{\sigma}\cdot{\bi e}_1} 
+ \frac{2}{R}\delta\gamma  =0,
\\
&&\dis{ {\bi e}_2\cdot \tensor{\sigma}\cdot{\bi e}_1} 
+  {\bi e}_2\cdot \nabla \delta \gamma ={\bi 0}, 
\ena  
in the normal and tangential directions, respectively.  
Here  $\delta\gamma=
\gamma -\gamma_0$  is the surface tension deviation and 
${\bi e}_2\cdot \nabla= R^{-1}\p/\p\theta$. 
From Eq.(2.25) the   tangential  gradient of the surface 
tension is equal to 
the  discontinuity of the tangential stress, 
which gives rise to a Marangoni flow \cite{Levich,Bedeaux1}.

As in the pure fluid case \cite{OnukiPRE}, 
we   assume that  the deviations of 
the temperature and the chemical potentials 
are continuous at the interface, 
\bea
\delta T&=&\delta T', 
\\
\delta \mu_1&=&\delta \mu_1' \quad 
\delta \mu_2=\delta \mu_2'.
\ena 
See Appendix B of our previous work \cite{OnukiPRE} 
and the item (i) in 
the summary section of this work 
for discussions 
on the validity of these assumptions.  
The Gibbs-Duhem 
relation for infinitesimal deviations  is written as  
$
(1-c)\delta \mu _1 +c \delta\mu_2 = -s\delta T+ 
\rho^{-1}\delta p,
$
which holds 
in  the liquid and gas regions close to the 
surface. Further use of Eqs.(2.26) and (2.27) yields     
\be 
\dis{c} \delta \mu
+\dis{s} \delta T= \frac{1}{\rho}\delta p-
\frac{1}{\rho'}\delta p', 
\en  
where $\delta \mu
=\delta\mu_2-\delta\mu_1$.

From Eq.(2.7) the mass 
conservation of the second  component 
at the surface gives  
\be 
\dis{c} J +
{\bi e}_1\cdot \dis{\bi I}=0.
\en 
From Eq.(2.11) the energy conservation at the interface gives 
\be 
 (\dis{s}T+  \dis{c}\mu) J+ {\bi e}_1
\cdot  \dis{\bi q}= 0,  
\en 
where use has been made of the thermodynamic relation 
$ 
e+p= \rho_1\mu_1+ \rho_2\mu_2+\rho sT
=(\mu_1+ c\mu +sT)\rho.
$
From Eqs.(2.29) and (2.30) $J$ may be removed to give 
\be 
{\bi e}_1
\cdot  \dis{{\bi q}-\alpha T {\bi I}}= 0,  
\en 
where the coefficient $\alpha$ is defined by  
\be 
\alpha= {\mu}/T + {[s]}/{[c]}.
\en 
Note that  $\alpha$ is a constant 
continuous across  the interface.
The flux ${\bi q}-\alpha T {\bi I}$ 
is continuous along the normal direction 
across the interface.

\subsection{Pressure and surface tension deviations}

We  shall see that  
the pressure deviations 
in the two phases are  negligibly small for 
large $R$ in Eq.(2.28).  This yields 
the following relation,    
\be 
\delta \mu  \cong -\frac{[s]}{[c]}\delta T =
\ppp{\mu}{T}{{\rm cx}, p}\delta T,
\en 
which  plays a key role in the following calculations. 
It may be  justified if the hydrodynamic deviations 
 are expanded in powers of the 
inverse bubble radius $R^{-1}$.  
To leading order in $R^{-1}$, the right hand side of Eq.(2.28) is 
 negligible, resulting in  Eq.(2.33).

In equilibrium,   the surface tension of binary mixtures 
is  defined on the coexistence surface in the 
space of three field variables such as $T$, $p$, and $\mu$.  
Here  the curvature effect is neglected for large $R$.  
Then  the surface tension is a function of 
 $T$ and $\mu$ if  $p$ is taken to be 
the coexistence pressure $p_{\rm cx}(T,\mu)$. In  Eq.(2.33) 
 the  temperature and chemical 
potential deviations near the interface 
are still on the coexistence surface in the isobaric condition. 
Thus  the surface tension   deviation is written  as \cite{note1}  
\bea 
\delta\gamma & = &  a_1 \delta T+a_2 \delta\mu\nonumber\\
&\cong& 
 -\gamma_1 \delta T,
\ena 
where $a_1$ and $a_2$ are the expansion coefficients.  
As discussed in Section 1, 
the coefficient $\gamma_1$ is written as  
\be 
-\gamma_1=
a_1 - a_2\frac{[s]}{[c]} 
=
\ppp{\gamma}{T}{{\rm cx},p}.
\en

\section{Axisymmetric solution}
\setcounter{equation}{0}

\subsection{Velocity and pressure}

In our problem,  the fluid flow is 
 axisymmetric with respect to the $z$ axis. 
The velocity field  ${\bi v} ({\bi r}) $ 
 and the pressure deviation $\delta p ({\bi r}) $
are  expressed in the same forms as in the one-component 
fluid \cite{OnukiPRE,Onukibook}.  That is, in terms of 
two functions   $\hat{Q}(r)$ 
and $\hat{ H}(r)$, ${\bi v}({\bi r})$ 
is written as 
\be
{\bi v} =
 \bigg (\frac{d\hat{H}}{dr}-\frac{\hat{H}}{r}+ 
\hat{Q}r\bigg) \frac{z}{r^2}{\bi r} 
+\frac{1}{r}\hat{H}{\bi e}_z. 
\en 
which satisfies Eq.(2.16). 
 Outside the bubble we have 
\be 
\hat{Q}={Q_1}\frac{R}{r^2}, \quad 
\hat{H}= \frac{R}{2}Q_1+ H_1\frac{R^3}{r^2} -v_{\rm D}r \quad (r>R). 
\en 
where $v_{\rm D}$ is the bubble velocity  
in the original reference frame. 
Inside the bubble we have  
\be 
\hat{Q}={Q_2'}\frac{r}{R^2}, \quad 
\hat{H}= -\frac{2}{5R^2}Q_2'r^3+ {H_2'}{r} \quad (r<R).    
\en 
Then Eq.(2.14) is satisfied.  In particular, Eq.(2.16) yields   
\be 
Q_1= gR^2(\rho-\rho')/3\eta.
\en 
The coefficients $Q_1$, $H_1$,  $Q_2'$, and $H_2'$ 
 have  the dimension of 
velocity.  From Eqs.(2.18) and (2.19) the  pressure 
deviation  is determined as  
\bea 
\delta p(r) &=& 
\eta Q_1 \frac{Rz}{r^3}- g\rho z \qquad\qquad (r>R) \nonumber\\ 
&=&   -2\eta' Q_2' \frac{z}{R^2} - g \rho' z\quad\quad (r<R).
\ena

The mass flux through the interface and the surface tensition deviation 
are  angle-dependent as 
\be 
J= J_1 \cos\theta,\quad \delta\gamma = \Gamma_{1} \cos\theta,
\en 
where $\cos\theta=z/r$ and  
$J_1$ and $\Gamma_{1}$  are constants. 
From the boundary conditions  Eq.(2.22)-(2.25),   
 $H_1$, $v_{\rm D}$,  $H_2'$, and $\Gamma_1$  may be expressed 
in terms of the three quantities $J_1$, $Q_2'$, and $Q_1$ as 
\bea 
H_1&=& \frac{1}{3}( \frac{1}{\rho'}-\frac{1}{\rho}) J_1 
-\frac{1}{15}Q_2' + \frac{1}{6}Q_1 ,
\\
v_{\rm D}&=& -\frac{1}{3}( \frac{2}{\rho'}+\frac{1}{\rho}) J_1 
+\frac{2}{15}Q_2' + \frac{2}{3}Q_1 ,
\\ 
H_2'&=&  \frac{1}{\rho'} J_1 + \frac{1}{5}Q_2' ,
\\ 
&&\hspace{-1cm}\Gamma_1= ( \frac{2}{\rho'}- \frac{2}{\rho})\eta J_1 
-(\frac{2}{5}\eta+\frac{3}{5}\eta') Q_2' 
+\eta Q_1 .
\ena 
Since $Q_1$ is determined as in Eq.(3.4), the two quantities 
 $J_1$ and $\Gamma_1$  remain unknown.

From Eqs.(2.18)  and (2.19) 
the deviation of the chemical potential difference 
$\delta\mu$ also satisfies $\nabla^2\delta\mu=0$ for $r\neq R$. 
Then $\delta T$ and $\delta\mu$ are   written as 
\bea 
\delta T ({\bi r}) &=& {({\cal T}' -{\cal T} ) }\frac{R^3}{r^3}z+ {\cal T} z 
\quad (r>R)
 \nonumber\\&=& {\cal T}'  z \qquad \qquad\qquad\quad  (r<R),
\\
\delta \mu ({\bi r}) &=& {({\cal M}' -{\cal M} ) }\frac{R^3}{r^3}z+ {\cal M} z 
\quad (r>R)
 \nonumber\\&=& {\cal M}'  z \qquad \qquad\qquad\quad  (r<R),
\ena 
where  $\cal T$ and $\cal M$ are the gradients far from 
the bubble defined in Eqs.(2.1) and (2.3), while 
${\cal  T}'$ and ${\cal  M}'$  are those  within it. 
At $r=R$,  $\delta T$ and $\delta \mu$ 
are continuous from  Eqs.(2.26) and (2.27). For 
$r\leq R$ we have   
\be 
\delta T=  {\cal T}'z, \quad 
\delta \mu=  {\cal M}'z.
\en 
In terms of ${\cal T}'$ and ${\cal M}'$ 
the relation Eq.(2.28) is rewritten as 
\be 
[c]{\cal M}'+ [s]{\cal T}'= \bigg[ \frac{\eta}{\rho}{Q_1} 
+\frac{2\eta'}{\rho'}{Q_2'}\bigg] \frac{1}{R^2} .
\en 
Furthermore, we may  derive two  equations for 
${\cal T}'$ and ${\cal M}'$ 
from Eqs.(2.29) and (2.30). Together with Eqs.(3.10)  and  (3.14), 
we have four equations, which   constitute a closed set 
 determining  $J_1$, $Q_2'$, ${\cal T}'$, 
and ${\cal M}'$.  Here, without assuming Eq.(2.33),    we may  assume 
the general relation  $\delta\gamma=a_1\delta T+a_2\delta \mu$ 
in the first line  of Eq.(2.34).

\subsection{Case of  $g\neq 0$ and ${\cal Q}={\cal I}=0$}

Here we consider the gravity-induced solution 
with  $g\neq 0$ and ${\cal Q}={\cal I}=0$, where  
all the coefficients in Eqs.(3.7)-(3.10) 
are proportional to $Q_1$ in Eq.(3.4).   If  
the pressure deviations in Eq.(2.28) are neglected, 
there arise no deviations of the temperature, 
the chemical potential,  
and the surface tension, 
 ${\cal T}={\cal T}'=J_1=\Gamma_1=0$.  in Eqs.(3.7)-(3.10). 
Then Eq.(3.10) gives 
\be 
 Q_2' ={5\eta Q_1}/{({2}\eta+{3}\eta')},
\en 
while 
Eq.(3.8) gives  $v_D=v_g$, where $v_g$ is  the well-known 
gravity-induced velocity \cite{Levich,Ha,Ry}.  
\be 
v_g=\frac{2(\eta+\eta')(\rho-\rho')}{3\eta(2\eta+3\eta')}R^2g.
\en  
In  the lowest order of $R^{-1}$ 
the right hand side of Eq.(2.28) at $r=R$ becomes   
\be 
 \frac{\delta p}{\rho}-
\frac{\delta p'}{\rho'}\cong \bigg[\frac{\eta}{\rho}+ 
\frac{10\eta\eta'}{\rho'(2\eta+3\eta')}\bigg] \frac{z}{R^2}Q_1,
\en
where the gravity terms cancel to vanish. 
If divided by $Q_1$, the above quantity is of order $R^{-1}$. 
If  we  assume 
the linear relation  $\delta\gamma=a_1\delta T+a_2\delta \mu$,  
 the above relation Eq.(3.17) leads to  
$\delta T \propto {\delta\mu}  \propto Q_1/R$, 
${\cal T}'  \propto{\cal M}'  \propto  Q_1/R^2$,  
 $\Gamma_1 \propto Q_1/R$, and 
$J_1 \propto Q_1/R^2$.   
Thus, in the presence of weak gravity only, 
 a large droplet or a large  bubble moves with 
the velocity $v_g$ in Eq.(3.16), where 
  first-order phase transition 
 and temperature inhomogeneities are negligible.

\subsection{Case of   ${\cal Q}\neq 0$, 
${\cal I}\neq 0$, and $g=0$}

We seek the solution in the presence of 
${\cal Q}$ and ${\cal I}$ in  
the gravity-free condition  $g=0$. 
Remarkably, $\Gamma_1\propto R$ in binary 
mixtures with phase change,  
while $\Gamma_1\propto R^{-1}$ in 
one-component fluids. 
We clarify the relationship 
of our theory and the previous theories: 
(i) To obtain the solution 
without phase change  \cite{Young}, we set  
$J_1=0$ and $\Gamma_1= -\gamma_1 R{\cal T}'\propto R$
 in Eqs.(3.7)-(3.10) and require 
the energy conservation relation 
${\bi e}_1\cdot[{\bi q}] =0$ from Eq.(2.30). 
(ii) To obtain the solution for 
one-component fluids with phase change \cite{OnukiPRE}, 
we neglect $\Gamma_1$ and retain $J_1$ 
in Eq.(3.10).

We  use the relation Eq.(2.33) or neglect the right hand side of 
Eq.(3.14) to  obtain 
\be 
{\cal M}'= -{\cal T}'[s]/[c].
\en 
The above relation will be  justified self-consistently 
at the end of this subsection.  
From Eqs.(2.34) and (3.13) 
 $\Gamma_1$ in Eqs.(3.6) and (3.10) is expressed as 
\be 
\Gamma_1= -\gamma_1 R {\cal T}'.
\en 
As will be shown in Appendix B, ${\cal T}'$ can be 
written as 
\be 
 {\cal  T}'=  \frac{3}{2\lambda_e+\lambda'_e}
({\cal Q}-\alpha T {\cal I}),
\en 
where ${\cal Q}$ and $\cal I$ are defined by 
Eqs.(2.14) and (2.15). We introduce 
 the effective thermal conductivity  by 
\be 
\lambda_e =\lambda+ \frac{\rho D}{T} \ppp{\mu}{c}{Tp} 
(k_T+Z)^2.
\en 
The $\lambda_e$ and $\lambda_e'$ in Eq.(3.20) are 
the values of $\lambda_e$ at $r=R+\pm 0$. 
In Eq.(3.21) we define  $Z$  by 
\be 
\frac{Z}{T}= 
 \ppp{c}{T}{\mu p}-\ppp{c}{\mu }{T p}\frac{[s]}{[c]} 
= \ppp{c}{T}{{\rm cx},p}.
\en 
On the other hand, the mass flux through the interface 
$J$ is calculated from Eq.(2.29) or Eq.(2.30). 
Then $J_1$ in Eq.(3.6) is expressed as  
\be
J_1= 3\frac{({2\lambda_a + \lambda_a' })
{\cal I} 
- ({2B + B'}){\cal Q}}{[c](2\lambda_e +\lambda_e')}, 
\en 
where  $\lambda_a$,  $\lambda_a'$, $B$, and $B'$  are the values of 
$\lambda_a$ and $B$ at $r=R+\pm 0$ with 
\bea 
\lambda_a &=&\lambda_e +\alpha T B=\lambda+AB,\\
B &=&\rho D (k_T+Z)/T.  
\ena

From Eq.(3.10) we obtain 
\be 
 Q_2' =  \frac{5}{{2}\eta+{3}\eta'} 
\bigg [ \gamma_1 R {\cal T}' 
+  (\frac{2}{\rho'} - \frac{2}{\rho}) \eta J_1 \bigg ].
\en 
Substitution of the above relation into  Eq.(3.8) yields   
 the drift velocity  composed of two parts,  
\be 
v_D= v_D^M+v_D^c,
\en 
where $v_D^M$ arises from the Marangoni 
effect and $v_D^c$ from the evaporation-condensation. 
They are written as 
\bea  
 v_D^M & =&  \frac{2\gamma_1}{3({2}\eta+{3}\eta')} 
  R{\cal T}',
\\
v_D^c& =&  \frac{-2}{{2}\eta+{3}\eta'} 
\bigg(\frac{\eta}{\rho}+\frac{\eta'}{2\rho}+
\frac{\eta'}{\rho'}\bigg)J_1  .
\ena
In terms of  these  characteristic velocities, 
the velocity field ${\bi v}({\bi r})$ 
in the reference frame moving with the bubble 
or the droplet is expressed as 
\be 
{\bi v}=  
v_D^M {\bi a}_M+  
\frac{\eta (\rho'-\rho) v_D^c}{\eta\rho' + \eta'(\rho+\rho'/2)}
 {\bi a}_c-(v_D^M+ v_D^c) {\bi e}_z,
\en  
where ${\bi a}_M={\bi a}_M({\bi r})$ 
and ${\bi a}_c={\bi a}_c({\bi r})$ 
are the following space-dependent dimensionless 
vectors,
\bea 
 {\bi a}_M&= & 
  -\frac{R^3}{2r^3}{\bi e}_z +\frac{3R^3}{2r^5}z{\bi r}
\quad (r>R) 
\nonumber\\
&=&   
\frac{5}{2} {\bi e}_z- \frac{3 r^2}{R^2}{\bi e}_z 
+\frac{3}{2R^2} z{\bi r} 
\quad (r<R), \\
{{\bi a}_c} 
&=& \frac{{\eta'}}{2\eta}\bigg [ 
\frac{R^3}{r^3}{\bi e}_z -\frac{3R^3}{r^5}z{\bi r}\bigg ]
\quad (r>R) 
\nonumber\\
&=&\frac{z{\bi r}}{R^2}+
\bigg[2+\frac{{\eta}'}{2\eta}-
\frac{2r^2}{R^2} \bigg] {\bi e}_z ~ (r<R),
\ena

We may now show that 
the right hand side of Eq.(3.14) is surely negligible for large $R$. 
Use of Eq.(3.26) gives  
\be 
\frac{\delta p}{\rho}-
\frac{\delta p'}{\rho'}= 
\frac{10\eta'}{{2}\eta+{3}\eta'}\bigg  
[ \frac{\gamma_1{\cal T}'}{\rho'R} 
+ (\frac{2}{\rho'}-\frac{2}{\rho}) 
 \frac{\eta J_1}{\rho' R^2}\bigg]  .
\en 
We  compare the term ($\propto R^{-1}{\cal T}'$) 
on the right hand side  of Eq.(3.33) and the term 
$[s]{\cal T}'$ on the left hand side 
of Eq.(3.14). The former is much 
smaller than the latter for $R\gg 
|\gamma_1/\rho'[s]|$, where 
the right hand side is microscopic. 
The term ($\propto R^{-2}J_1$) due to 
the evaporation-condensation 
in Eq.(3.33) is also negligible, as already 
verified in our previous paper \cite{OnukiPRE}.

The concentration deviation 
$\delta c({\bi r})$ can also be expressed in the same 
form as in Eqs.(3.11) and (3.12), where   
the gradient $\cal C$ far from the droplet  satisfies Eq.(2.3), 
Its values at $r=R\pm 0$ are written as $\delta c_{\pm}$ 
and are  expressed as 
\be 
\delta c_{+}= \frac{Z}{T}{\cal T}'z,
\quad 
\delta c_{-}= \frac{Z'}{T}{\cal T}'z,
\en 
where $Z$ and $Z'$ are the values 
of $Z$ in Eq.(3.22) at $r=R\pm 0$. 
If  $Z\neq Z'$, $\delta c$ is  
 discontinuous at $r=R$.


We  note that  the Marangoni effect vanishes  in  
 azeotropic mixtures \cite{Onukibook}, where 
 the two phases 
have the same composition or $[c]=0$. 
There is also no difference 
in the molar fractions of the 
two phases (see  the sentence below Eq.(C3)). 
Special analysis is thus needed when we treat 
nearly azeotropic mixtures. For example, 
in H$_2$O-D$_2$O mixtures, 
 the relative composition change 
$[c]/c$ is only $0.5\%$ of the relative 
density change  $[n]/n$ near the critical line \cite{Russia}. 
Thus let us consider the limit  $[c]\to 0$ 
in the equations 
in this subsection. Then 
 $Z \sim [c]^{-1}$ and $\lambda_e \sim  [c]^{-2}$ from Eqs.(3.20) 
and (3.21), leading to 
$\Gamma_1 \sim \gamma_1 [c]^2$, 
 ${\cal T}' \sim [c]^2 ({\cal Q}- \alpha T {\cal I})$, and 
\be 
v_D^M \sim \gamma_1 [c]^2R ({\cal Q}- \alpha T {\cal I}). 
\en 
from Eqs.(3.19), (3.20), and (3.28). Here $\gamma_1 \sim [c]^{-1}$ 
\cite{OnukiJCP} and $\gamma_1[c]^2  \sim [c]$, 
so the Marangoni flow is of order  $[c]$.

\section{Dilute mixtures in gas-liquid 
coexistence }
\setcounter{equation}{0}

Let the second component be  a dilute solute. 
Under the condition Eq.(A8) we set 
\bea 
&&c\cong m_2 X/m_1, \\ 
&&(\p c/\p \mu)_{Tp}\cong 
{m_2^2X}/{m_1 T},
\ena 
where $X$ is the molar fraction 
and $m_1$ and $m_2$ are the molecular masses. 
The second relation (4.2) does not hold very close to the 
solvent criticality even for small $X$ (see Eq.(C7)).  
Hereafter the Boltzmann constant will be set equal to 
unity.  In the literature 
\cite{OnukiJCP,Sengers1,De,Anisimov,Shock,Russia},  
Henry's law is expressed in terms of the solute  molar 
fractions.  That is, in equilibrium, 
the solute molar fraction in gas $X_g$ and 
that in liquid $X_\ell$ are related by 
the partition coefficient  \cite{OnukiJCP}, 
\be 
{\cal K}=X_g/X_\ell,
\en 
which depends  on $T$ along the solvent coexistence 
line $p=p_{\rm cx}(T)$. 
In dilute mixtures $[c]/c=[X]/X$ holds and 
 $[c]/c$ is independent of $c$ as       
\bea 
[c]/c  &=&1-{\cal K}  \quad ({\rm gas~ bubble})\nonumber\\
&=& 1- {\cal K}^{-1} \quad ({\rm liquid~ droplet}).  
\ena

\subsection{Expressions as  $c \to 0$ }

To simplify the notation 
we introduce the following dimensionless parameter,
\be 
{\cal W}= \frac{[X]}{X[\sigma]}= \frac{[c]}{cm_1[s]},
\en 
where $\sigma=m_1s$ is the entropy per solvent particle. 
This parameter tends to a well-defined limit 
in the dilute limit, vanishes for  azeotropic mixtures, 
and  becomes proportional  to   the Krichevskii parameter  
 $K_{\rm Kr}$ near the solvent criticality as in 
Eq.(C6) in Appendix C.  
In  Eq.(3.22) $Z$ behaves  in terms of ${\cal W}$ as 
\be 
Z\cong -\frac{m_2c[s]}{[c]}= - \frac{m_2}{m_1{\cal W}}.
\en  
From Eq.(3.21) $\lambda_e$ 
is inversely proportional to $c$ as 
\be 
\lambda_e\cong \frac{\rho D_0}{m_2c} 
 Z^2\cong  \frac{ n D_0}{ {\cal W}^2X},
\en 
where $D_0=\lim_{c\to 0}D$ is the diffusion constant 
of a single solute molecule and $n=\rho/m_1$ is the 
solvent number density. 
The thermal conductivity $\lambda$  in Eq.(3.21) 
is smaller than the right hand side of 
 Eq.(4.7), as will be discussed in  Appendix C. 
Here, in calculating the flux 
${\bi e}_1\cdot({\bi q}- \alpha T {\bi I})$ 
in Eq.(2.31),  we have picked  up 
 the contribution from the solute diffusion 
to obtain  $\lambda_e$ in Eq.(4.7).

Furthermore,  $\lambda_a$  in Eq.(3.24) and  $B$ in Eq.(3.25) 
are finite as $c\to 0$. From Eq.(3.20) 
 we thus find ${\cal T}'\propto c$  as 
\be 
{\cal T}' 
\cong
A_M c \bigg({\cal Q}- T\frac{[s]}{[c]} {\cal I}\bigg).   
\en  
Here  we have set $\alpha \cong [s]/[c]$ from Eq.(2.32) 
assuming that $\cal I$ is of order $c$. 
The coefficient $A_M$ is a positive constant 
independent of $c$ as $c\to 0$ and is defined by 
\be 
A_M = \frac{3{\cal W}^2m_1/m_2}{{2n D_0+ X'n' D_0'/X}},
\en  
which is expressed  in terms of the molar 
fractions $X$ and $X'$  
outside and inside  the domain. 
The mass flux  through the interface 
$J$ in Eq.(2.22) 
tends to that of the pure fluid as $c\to 0$. That is, 
$J_1$ in Eq.(3.6) becomes  
\be 
J_1= (3/\dis{s}T)  {\cal Q}.  
\en 
Now the two drift velocities $v_D^M$ in Eq.(3.28) and 
$v_D^c$  in Eq.(3.29) are written as 
\bea  
 v_D^M & =&  \frac{2A_M\gamma_1}{3({2}\eta+{3}\eta')} 
 c R\bigg  ( {\cal Q}- \frac{[s]}{[c]} T {\cal I}
\bigg),
\\
v_D^c& =&  \frac{-6}{{2}\eta+{3}\eta'} 
\bigg(\frac{\eta}{\rho}+\frac{\eta'}{2\rho}+
\frac{\eta'}{\rho'}\bigg)
\frac{{\cal Q}}{[s]T}  .
\ena

From Eq.(3.34) the deviation of the 
mass fraction  
$\delta c$ at $r=R\pm 0$ are expressed as 
\be 
\frac{\delta c_{+}}{c}=
\frac{\delta c_{-}}{c'}= 
  - \frac{m_2[s]}{[c]}{\cal T}'z.
\en
Thus  $\delta c/c$ is 
continuous and 
$[\delta c]= -m_2[s]{\delta  T}$ at $r=R$ 
to leading order in $R^{-1}$. 
In our linear theory we require $
|\delta c|\ll c$, which becomes 
$|{\cal T}'|R\ll |[c]/m_2[s]|$. 
From Eq.(4.8) this inequality 
is satisfied for small 
$\cal Q$ and $\cal I$ even as $c\to 0$.

Let us consider situations 
in which the diffusion flux $\cal I$ is negligible 
in $v_D^M$ in Eq.(4.11). 
Then  
 $v_D^M$ and $v_D^c$ 
 have the same  sign  for  
a bubble with $\gamma_1>0$ and for a liquid droplet 
with $\gamma_1<0$, 
while they  have different signs for  
a bubble with  
$\gamma_1<0$ and for a liquid droplet 
with $\gamma_1>0$. In accord with the experiments 
\cite{Abe,Savino,Savino1},  
bubbles  can move towards cooler regions 
 with increasing the concentration of 
 a solute in the case 
$\gamma_1<0$.

\subsection{Dilute mixtures near the solvent criticality}

The above expressions can be used even 
in the vicinity of the   solvent criticality 
under the condition $X\ll n/{\cal W}^2 C_p$ in 
Eq.(C8), where $C_p$ is the isobaric specific heat per unit volume. 
In the near-critical case, the reduced temperature, 
\be 
\epsilon= 1-T/T_c,
\en 
is small (say, less than $10^{-3}$) and 
the differences between the two phases 
tend to vanish, so  $\rho'\cong \rho$ and  
$\eta'\cong \eta$.  In  Eq.(4.9) we have 
\be 
A_M \cong {\cal W}^2m_1/m_2nD_0.
\en 
Defining the solute 
hydrodynamic radius $a_0$ 
using  the Stokes formula $D_0 =T/6\pi \eta a_0$, we obtain 
\bea  
 v_D^M & =&  \frac{4\pi}{5nT} {\cal W}^2\gamma_1a_0RX 
\bigg 
 ( {\cal Q}- \frac{[s]}{[c]} T {\cal I}
\bigg),
\\
v_D^c& =&  {-3} {{\cal Q}} /{nT[\sigma]} ,
\ena
where $[\sigma]=m_1[s] \sim \epsilon^\beta$.

With increasing $X$ from zero, crossover occurs from 
the pure-fluid behavior to the mixture-behavior for 
$X>X^*$, where the crossover molar fraction is   
\be 
X^*=(\epsilon^\beta |\gamma_1| {\cal W}^2 a_0)^{-1}  R^{-1}.
\en
This is equivalent  to Eq.(1.6) near the solvent criticality 
with $c^*=m_2X^*/m_1\sim X^*$ 
if $|{\cal W}|\sim 1$ and $m_2\sim m_1$.  
The right hand side is of order 
$\epsilon^{1-\nu-\beta}\xi/R$ for 
$|{\cal W}|\sim 1$ and $\gamma_1 \sim -d\gamma_0/dT$ with  
$\nu\cong 0.625$ and $\beta\cong 0.33$,  
$\xi (\sim \epsilon^{-\nu})$ being  the correlation 
length ($\sim$the interface thickness).

In near-critical 
pure fluids in the gravity-free condition, 
a bubble in liquid 
was observed to be attracted 
to a warmer boundary     
\cite{Beysens}.  With addition of a small 
amout of various   solutes, 
it is then of great interest 
whether  a bubble is more  
attracted to or eventually repelled 
from a warmer boundary.   Here 
the  crossover concentration  $X^*$  
should  be measured  to confirm the theoretical 
expression (4.18).

\begin{figure*}[t]
\includegraphics[scale=0.36]{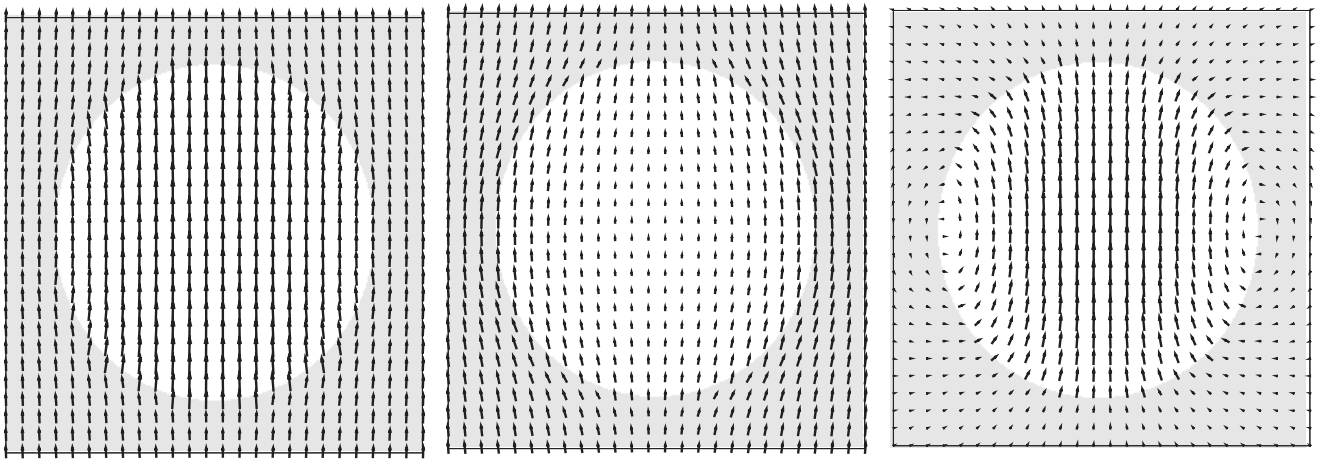}
\caption{\protect
Gravity-free  velocity field  around a bubble 
for $\eta'/\eta=\rho'/\rho=0.5$ in the $x$-$z$ plane 
in Eqs.(3.30)-(3.32), 
where   $v_D^M=0$ (left), 
$v_D^M=v_D^c$ (middle), and 
 $v_D^M=-v_D^c$ (right).}
\end{figure*}
-

\begin{figure*}[t]
\includegraphics[scale=0.36]{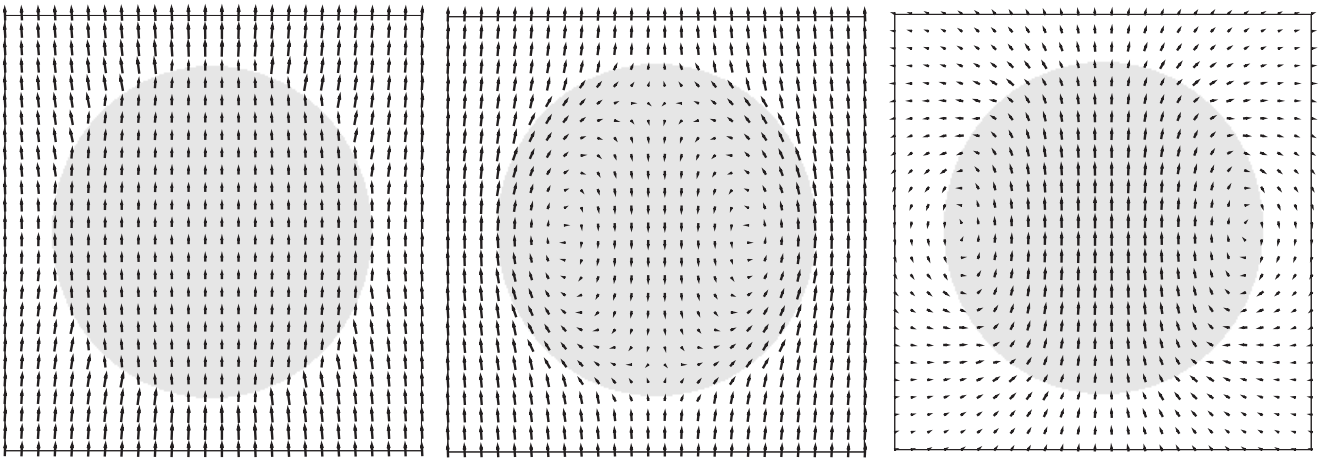}
\caption{\protect
Gravity-free  velocity field  around a liquid droplet  
for $\eta'/\eta=\rho'/\rho=2$ in the $x$-$z$ plane in Eqs.(3.30)-(3.32), 
where   $v_D^M=0$ (left), 
$v_D^M=v_D^c$ (middle), and 
 $v_D^M=-v_D^c$ (right).}
\end{figure*}

\section{Velocity profiles for  ${\cal Q}\neq 0$, 
${\cal I}\neq 0$, and $g=0$}

Young {\it et al.}  \cite{Young} calculated 
the  velocity field 
 without phase change due to 
the Marangoni effect  for $\gamma_1>0$ 
(which is given by  Eq.(3.30) with $v_D^c=0$). 
See its dipolar profile in 
their paper and in our previous paper \cite{OnukiPRE}.  
On the other hand, our previous paper \cite{OnukiPRE} 
has presented some examples of the velocity field 
due to  first-order phase change   
for pure fluids (which is given by  Eq.(3.30) with $v_D^M=0$). 
Here we display the velocity fields 
realized in dilute mixtures with increasing $c$ or 
in nearly azeotropic mixtures.

Without gravity, we show  
the velocity field $(v_x,v_z)$ 
in the $x$-$z$ plane 
for  a bubble with $\rho'/\rho=\eta'/\eta=0.5$ 
in Fig.1 and for a liquid droplet with $\rho'/\rho=\eta'/\eta=2$ 
in Fig.2.  It is written in the reference frame 
moving with a  bubble or a   droplet  
in the case $v_D^c<0$. 
Here the liquid density is twice larger 
than the gas density, which is realized for  
$T/T_c=0.97$ in the van der Waals theory of pure fluids. 
In the left panels,  
we set $v_D^M =0$ for pure fluids or for $\gamma_1=0$ 
in binary mixtures. 
In  the middle panels, we set 
 $v_D^M =v_D^c$, where 
the two mechanisms equally contribute 
to the drift velocity.  
In  the right  panels, we set 
 $v_D^M =-v_D^c$, where 
the drift velocity $v_D$ in Eq.(3.27) vanishes 
and the bubble or the droplet 
is at rest.

\section{Summary and remarks}
\setcounter{equation}{0} 

We have examined the competition 
of the evaporation-condensation effect 
and the Marangoni effect in the 
motion of a bubble 
or a droplet in weak heat and diffusion 
fluxes in binary mixtures. 
We have treated the simplest case of 
steady states with a constant drift velocity 
in the axisymmetric geometry, though 
the nonlinear terms in the hydrodynamic equations 
cannot be neglected in practical applications using large 
bubbles.   In  non-azeotropic 
binary mixtures, the crossover 
occurs from the evaporation-condensation mechanism  
 to the Marangoni mechanism 
  at a very small 
solute concentration. 
In our theory,  the coefficient 
$\gamma_1$ in Eq.(1.3) controls the strength of the 
Marangoni flow in heat flux,  
which can be both positive and negative 
depending on the solvent 
and solute species \cite{OnukiJCP}. 
The Marangoni flow is induced in opposite directions 
in the normal case $\gamma_1>0$ 
and in the anomalous case  $\gamma_1<0$, though 
the case $\gamma_1>0$  has mostly been studied.

Some further remarks are given below.\\
(i) We have assumed the continuity of the temperature 
and the chemical potentials 
and neglected the pressure deviations at the interface 
to obtain the key relation Eq.(2.33) or Eq.(3.18). 
It means that  the interface stays on the 
coexistence surface $p=p_{\rm cx}(T,\mu)$ 
even in nonequilibrium. This is justified for large $R$. 
However, if the gas phase is very dilute far below 
the critical temperature,  the surface dissipation 
mainly occurs in the gas phase side within a distance of the 
mean-free path  inversely proportional to the gas  density
\cite{Pao,Sone,Ward}. 
There can then be an apparent 
 temperature jump at the interface.\\ 
(ii) The behavior of the coefficient $\gamma_1$ in Eq.(1.3)  
or in Eq.(2.35) is highly nontrivial. 
Theoretically, it has been 
examined  only for dilute mixtures \cite{OnukiJCP}.
Its behavior is also  of interest for  
binary mixtures near 
a lower critical solution temperature 
 (LCST).  In  a phase-separated mixture of 
butoxyethanol-water near its LCST,  
Braun {\it et al.} \cite{space} applied heat-pulses 
to water-rich droplets to observe  
their motion from a high-temperature 
region to a  low-temperature region. 
This motion  was due to the Marangoni effect 
because the motion was  in the direction of 
decreasing the surface tension. 
\\
(iii)  Boiling on a heated substrate has been of great 
interest  both on earth  and in space 
\cite{Straub}. The effect of  
a noncondensable gas should be studied in future. 
In accord with this paper, Marek and Starub  
\cite{St} claimed that 
the temperature gradients along the bubble interface 
inducing a Marangoni flow 
are caused by saturation pressure gradients 
due to a nonuniform accumulation of a  noncondensable gas
along the interface. Such flow serves to 
suppress detachment of bubbles 
 for $\gamma_1>0$, but should accelerate it 
 for  $\gamma_1<0$ \cite{Abe}.
\\
(iv) In gravity-free conditions, 
a spherical bubble or droplet can be 
suspended in liquid or gas 
in equilibrium. 
It is   of great interest 
how the velocity field and temperature 
evolve  after application 
of heat flux from a boundary. 
The  piston effect  comes into play on 
acoustic  time scales 
 \cite{Onukibook,Ferrell}.  
On longer time scales, a small amount of 
a solute should  drastically change the 
hydrodynamic behavior inducing a Marangoni flow. 
However, since we have treated only steady  states,  
it remains unclear how 
the concentration changes in time along the interface.\\ 
(v) Thermocapillary hydrodynamics 
 has been  puzzling  
 near the critical point 
\cite{Beysens,Onukibook,Ferrell}, where  
the singularities of  the thermodynamic and dynamical 
properties largely influence the dynamics.  
The condition (4.2) or the condition (C8) 
does not  hold sufficiently 
close to the critical point, 
where the results in Section 4 
cannot be used.  Thus, 
with addition of a solute, 
 two-phase hydrodynamics 
poses a  new problem of critical dynamics. \\
(vi)  Finally, we 
should stress that  
surfactant molecules 
absorbed at interfaces 
give rise to Marangoni flow \cite{Levich,Onuki1993}, though  
this effect is beyond the scope of this paper.  
\\

\vspace{3mm}

{\bf Acknowledgments}\\
This work was supported by Grants-in-Aid 
for scientific research 
on Priority Area ``Soft Matter Physics" 
and  the Global COE program 
``The Next Generation of Physics, Spun from Universality and Emergence" 
of Kyoto University 
 from the Ministry of Education, 
Culture, Sports, Science and Technology of Japan.

\vspace{2mm} 
{\bf Appendix A: Onsager relations}\\
\setcounter{equation}{0}
\renewcommand{\theequation}{A\arabic{equation}}

In  the linear response theory,  
the dissipative heat and 
diffusion  fluxes, $\bi q$ and $\bi I$, in 
 binary mixtures  are expressed in terms of  the  
Onsager kinetic  coefficients 
$L_{ij}$ \cite{Landau}. The thermodynamic forces 
are $\nabla T$ and $\nabla(\mu/T)$ as  
\bea 
{\bi q}&=& -  L_{11} \nabla T - 
TL_{12} \nabla \frac{\mu}{T} , \\
{\bi I}&=& - L_{12}\frac{1}{T}\nabla T -L_{22} \nabla \frac{\mu}{T},
\ena  
where $\mu={\mu_2-\mu_1}$ with $\mu_i$ being the 
chemical potentials per unit mass.  We then consider 
the dynamic equation for the 
entropy density $S=\rho s$ per unit volume.  
The  thermodynamic relation 
$TdS  = de-\mu_1d\rho_1-\mu_2d\rho_2$ 
and the hydrodynamic equations yield  
\be 
\frac{\p }{\p t}S+ \nabla\cdot\bigg(S{\bi v}+ \frac{1}{T}{\bi q} 
-\frac{\mu}{T} {\bi I}\bigg) = \frac{1}{T}\dot{\epsilon} , 
\en 
where $\dot\epsilon$ is 
 the heat production rate per unit volume  expressed as 
\be 
\dot{\epsilon}=\sum_{ij}{\sigma_{ij}}\frac{\p v_i}{\p x_j} 
 - {{\bi q}} \cdot\frac{\nabla T}{T}
- T{\bi I}\cdot\nabla \frac{\mu}{T}. 
\en 
The first term arises from the viscous damping 
and the last two terms from the heat conduction and diffusion. 
As is well-known,  $\dot\sigma$ is nonnegative-definite 
if the coefficients 
$L_{ij}$ constitute a symmetric positive-definite $2\times 2$ 
matrix.  It is well-known that $L_{ij}$ are 
expressed in terms of the the time-integral 
of the appropriate flux time-correlations (Green-Kubo 
formulas) \cite{Onukibook}.

It is convenient to express $\bi q$ as in Eq.(2.12) 
and $\bi I$ as in Eq.(2.8). Then 
$\lambda$, $A$, $D$, and $k_T$ are expressed as 
\bea 
&&\lambda = L_{11}-L_{12}^2/L_{22},\\
&&A=TL_{12}/L_{22} ,\\
&&\rho T D= {L_{22}}  \ppp{\mu}{c}{p T},\\
&& \rho D k_T=L_{12} -\frac{\mu}{T}L_{22} + \ppp{\mu}{T}{c p }L_{22} .
\ena 
If  $L_{12}$ and $L_{22}$ are removed from Eqs.(A6)-(A8),  
  $k_T$  and  $A$  are related as 
in Eq.(2.13).

\vspace{2mm} 
{\bf Appendix B: Calculations of ${\cal T}'$ 
and $J_1$  }\\
\setcounter{equation}{0}
\renewcommand{\theequation}{B\arabic{equation}}

To derive  Eq.(3.20) we calculate  
 the heat and diffusion fluxes 
at the interface in the normal direction 
substituting Eqs.(3.11) and (3.12) into  Eqs.(A1) and (A2) 
and setting ${\cal M}'= -{\cal T}'[s]/[c]$ from Eq.(3.14) 
for large $R$.  
The unprimed quantities are  the values  at $r=R+0$, 
while the primed ones are those at $r=R-0$.

From Eq.(3.11) 
the gradient  ${\bi e}_1\cdot\nabla\delta T$ normal to the 
interface is  $(3{\cal T}-2{\cal  T}')\cos\theta$ for $r=R+0$ 
and to  ${\cal  T}'\cos\theta$ for $r=R-0$. 
From Eq.(3.12) the gradient  ${\bi e}_1\cdot\nabla\delta \mu$ 
is obtained  by replacement of ${\cal  T}'$ and ${\cal  T}$ by 
 ${\cal  M}'$ and ${\cal  M}$. 
Then  use of Eqs.(3.11)-(3.14) gives  
\bea 
{\bi e}_1\cdot{\bi q} &=&  
(2\lambda_a {\cal T}'- 3{\cal Q})\hat{z} 
\quad (r=R+0),\nonumber\\
&=&  
-\lambda_a' {\cal T}'\hat{z}  
\quad (r=R-0),
\\
{\bi e}_1\cdot{\bi I} &=&  
( 2B {\cal T}'-3{\cal I})\hat{z}  
\quad (r=R+0),\nonumber\\
&=&  
-B' {\cal T}'\hat{z} 
\quad (r=R-0)
\ena 
where ${\hat z}=z/r=\cos\theta$. 
We define 
$\cal Q$ and $\cal I$ in  Eqs.(2.14) and (2.15) 
and introduce  
\bea 
\lambda_a &=&L_{11}-
\alpha L_{12}=\lambda +L_{12} (A/T-\alpha),
\\ 
B &=&(L_{12}-\alpha L_{22})/T= L_{22}(A/T-\alpha)/T.
\ena  
The flux ${\bi q}-T\alpha{\bi I}$ along ${\bi e}_1$ 
in Eq.(2.31) 
may then be calculated at $r=R\pm 0$. 
Its continuity at $r=R$ gives  
\be  
(2\lambda_e+\lambda_e') {\cal T}'- 3({\cal Q}-T\alpha{\cal I})=0 .
\en 
Here $\lambda_e$ is the effective thermal conductivity   
defined by   
\bea 
\lambda_e &=& \lambda_a-\alpha BT \nonumber\\
&=&\lambda + L_{22} (A/T -\alpha)^2,
\ena 
where use of Eqs.(2.13) and (2.32) yields 
\bea 
\frac{A}{T}-\alpha &=& \ppp{\mu}{c}{Tp}\frac{k_T}{T}
- \ppp{\mu}{T}{cp}- \frac{[s]}{[c]}\nonumber\\
&=& \frac{1}{T}\ppp{\mu}{c}{Tp}({k_T+Z}). 
\ena 
From   $(\p c/\mu)_{Tp}(\p \mu/\p T)_{cp}=
-(\p c/\p T)_{\mu p}$ and Eq.(A7) 
we obtain  $Z$ in Eq.(3.22) and $\lambda_e$ in Eq.(3.21). 
From Eqs.(2.29) and (3.6) $J_1$ in the mass flux through 
the interface is determined by  
\bea
J_1&=& 
[3{\cal Q}-(2\lambda_a+\lambda_a'){\cal T}']/T[c]\alpha
\nonumber\\
&=&[ 3{\cal I}- 
(2B + B' ){\cal T}']/[c], 
\ena
which yields  Eq.(3.23) with the aid of 
Eq.(3.20).

\vspace{2mm} 
{\bf Appendix C: Mass and molar fractions 
and relations in near-critical dilute mixtures}\\
\setcounter{equation}{0}
\renewcommand{\theequation}{C\arabic{equation}}

In hydrodynamic theory the mass fraction $c$ and the 
chemical potential difference $\mu$ per unit mass 
are  usually used, but in thermodynamics it is 
convenient to use the molar fraction $X$ and the 
chemical potential difference $\Delta$ per particle. 
In terms of the molecular masses $m_1$ and $m_2$ 
they are related by \cite{Onukibook}
\bea 
c&=&{m_2 X}/[{m_1(1-X)+m_2X}],\\
\Delta&=&  m_2\mu_2-m_1 \mu_1. 
\ena 
At any concentration it generally holds  the relation,
\be 
\ppp{c}{\mu}{TP} =
 m_1^2m_2^2 (n/\rho)^3 \ppp{X}{\Delta}{TP}.
\en
We also note the relation $[c]= [X](m_1c c'/m_2 XX')$. 
The azeotropy condition 
 $[X]=0$ is also given by $[c]=0$.

First we give thermodynamic 
relations in dilute mixtures with $c\ll 1$. 
In the text of this paper we assume Eqs.(4.1) and (4.2). 
To describe the critical behavior, 
it is convenient to introduce  
the  Krichevskii parameter $K_{\rm Kr}$, which is 
the dilute limit of the thermodynamic derivative 
$({\p p}/{\p X})_{n T}$ at fixed $n=n_1+n_2$  
and $T$ at the solvent criticality 
\cite{Sengers1,De,Anisimov,Shock,Russia}.  
It is related to the 
derivatives of the critical pressure $p_c$ 
and the critical temperature $T_c$ with respect to $X$ 
along the critical line as 
\be 
K_{\rm Kr}= \frac{dp_c}{dX} - p'_{\rm cx} \frac{dT_c}{dX},
\en 
where $p'_{\rm cx}=dp_{\rm cx}/dT$ is the temperature-derivative of the 
coexistence pressure $p_{\rm cx}(T)$ of the pure fluid. 
The Clausius-Clapeyron relation 
$p'_{\rm cx} =[\sigma]/[v]$ holds for the pure fluid, 
where $v=1/n$ is the inverse density and $\sigma$ 
is the entropy per particle.  
In near-critical 
two-phase coexistence,  the mass fraction  difference 
$[X]$ and the volume difference $[v]$ are 
related by \cite{OnukiJCP}  
\be 
{[ X]}/{[ v]} =  ({K_{\rm Kr}}/{T_{c}}) X_c, 
\en 
where   $X_c 
=(X_g+X_\ell)/2$ is the critical value of $X$ \cite{OnukiJCP}. 
Thus the parameter $\cal W$ 
in Eq.(4.5) is expressed near the criticality as 
\be 
{\cal W}= {K_{\rm Kr}}[v]/{T_{c}}[\sigma]=  
K_{\rm Kr}/T_c p'_{\rm cx}.
\en
On the other hand,  the 
the thermodynamic derivative $(\p X/\p \Delta)_{Tp}$ 
behaves as \cite{Onukibook}
\be 
\ppp{X}{\Delta}{pT}\cong 
\frac{X}{T}  + \frac{X^2}{nT^2} K_{\rm cr}^2  K_{T\Delta}. 
\en 
where $K_{T\Delta}= (\p n/\p p)_{T\Delta}/n$ 
is the isothermal compressibility 
growing strongly near the criticality. 
On the right hand side of Eq.(C7), 
the first term is the dilute limit, while 
 the second term 
 is the singular contribution 
stemming   from the solute-solvent interaction and 
can be important  
very close to the criticality. 
In this paper we neglect the second term in Eq.(C7), 
which is allowable under  
the condition,  
\be 
{X}  \ll \frac{nT}{K_{\rm cr}^2 K_{T}}\cong \frac{n}{{\cal W}^2 C_{p}},
\en 
where $K_T=(\p n/\p p)_T/n$ 
is the isothermal compressibility and 
$C_p=\rho T(\p s/\p T)_p\cong T (p'_{\rm cx})^2K_T$ 
is the isobaric heat capacity per unit volume 
of the pure fluid \cite{Onukibook}. However, 
 the reverse relation 
$X > {n}/{{\cal W}^2 C_{p}}$  eventually  
holds sufficiently close to the criticality.

Next we consider the dynamic properties of dilute mixtures. 
The Onsager coefficients 
 $L_{12}$ and $L_{22}$ are proportional 
to $c$, while 
$L_{11}$ tends to the thermal conductivity 
of the pure fluid.   Thus, as $c\to 0$, 
  $A$ in Eq.(A6) and $D$ in Eq.(A7)  tend
 to well-defined limits, while 
$k_T$ in Eq.(A8) is proportional to $c$  as  
\be 
k_T= k_T^*c, 
\en 
with $k_T^*$ being a constant independent of $c$. 
The singular part of 
$L_{22}$ (proportional to $X^2$) is negligible 
compared to the background part under 
Eq.(C8) \cite{Onukibook}, so that   
\be 
L_{22}\cong  m_2\rho  c D_0,   
\en  
where $D_0$ is the 
diffusion constant of a single solute particle 
in the dilute limit.
The thermal conductivity 
$\lambda$ in Eq.(A5) 
behaves in  a dilute binary mixture  
near the solvent criticality  as \cite{Onukibook} 
\be 
\lambda\cong \lambda_s\lambda_B/(\lambda_s+\lambda_B) .
\en 
Here $\lambda_s$ is the thermal conductivity of the pure fluid 
growing strongly near the solvent criticality 
and $\lambda_B= \lambda_{B0}/X$ 
is the critical value inversely 
 proportional to $c$ with $\lambda_{B0}$ being a constant. 
The mode-coupling theory of critical dynamics 
predicted  the growing behavior 
$\lambda_s \cong TC_p/6\pi \eta \xi \sim \xi$, 
where $\xi$ is the correlation length and 
$C_p (\sim K_T)$ is the isobaric heat capacity per unit volume. 
We then recognize  that the effective 
thermal conductivity $\lambda_e$ in Eq.(3.21) 
may be approximated as 
 Eq.(4.6) under Eq.(C8). 
Using Eq.(C8) and the 
hydrodynamic radius $a_0$ 
in the Stokes formula $D_0=T/6\pi\eta a_0$,  we can make 
the following estimation, 
\be 
\frac{\lambda_e}{\lambda} >  
\frac{\lambda_e}{\lambda_s} \sim 
\frac{nD_0}{{\cal W}^2X\lambda_s}\sim 
\frac{n\xi/a_0}{C_p {\cal W}^2X}\gg 1.
\en


\begin{references}
\bibitem{note1} In dilute mixtures 
the expansion with respect to 
the   fugacity $f_2=e^{\mu/T}$ is 
more convenient, since $f_2$ tends to zero 
as $c\to 0$ \cite{OnukiJCP}.
\end{references}
\end{document}